\documentclass[a4paper,fleqn,usenatbib,article]{mnras}
\usepackage{newtxtext,newtxmath}
\usepackage[T1]{fontenc}
\usepackage{ae,aecompl}

\usepackage{graphicx}	

\usepackage{gensymb}
\usepackage{amsmath}	

\usepackage{threeparttable}
\usepackage{xcolor} 
\usepackage[normalem]{ulem}

\def\kpc{{\rm\,kpc}}
\def\kms{{\rm\,km\,s^{-1}}}
\def\msun{{\rm\,M_\odot}}

\def\pc{{\rm\,pc}}
\def\Mpc{{\rm\,Mpc}}

\def\eg{{e.g.,\ }}
\def\ie{{i.e.,\ }}

\def\lta{\mathrel{\spose{\lower 3pt\hbox{$\mathchar"218$}}
     \raise 2.0pt\hbox{$\mathchar"13C$}}}
\def\gta{\mathrel{\spose{\lower 3pt\hbox{$\mathchar"218$}}
     \raise 2.0pt\hbox{$\mathchar"13E$}}}
     
 \def\Gyr{{\rm\,Gyr}}

\def\FeH{{\rm[Fe/H]}}
\def\alfe{{\rm[\alpha/Fe]}}

\definecolor{forestgreen}{rgb}{0.13, 0.55, 0.13}

\title[Exploring the origin of low-metallicity stars]{Exploring the origin of low-metallicity stars in Milky Way-like galaxies with the NIHAO-UHD simulations}

\author[Federico Sestito et al.]{Federico Sestito,$^{1}$\thanks{E-mail: federico.sestito@astro.unistra.fr}
Tobias Buck,$^{2}$ Else Starkenburg,$^{2,3}$ Nicolas F. Martin,$^{1,4}$ 
\newauthor Julio F. Navarro,$^{5}$ Kim A. Venn,$^{5}$ Aura Obreja,$^{6}$ Pascale Jablonka,$^{7,8}$
\newauthor \& Andrea V. Macci\`o$^{9,10,4}$
\\
$^{1}$ Universit\'e de Strasbourg, CNRS, Observatoire astronomique de Strasbourg, UMR 7550, F-67000, France\\
$^{2}$ Leibniz Institute for Astrophysics Potsdam (AIP), An der Sternwarte 16, D-14482 Potsdam, Germany\\
$^{3}$ Kapteyn Astronomical Institute, University of Groningen, Landleven 12, 9747 AD Groningen, The Netherlands\\
$^{4}$  Max-Planck-Institut f\"ur Astronomie, K\"onigstuhl 17, D-69117, Heidelberg, Germany\\
$^{5}$ Department of Physics and Astronomy, University of Victoria, PO Box 3055, STN CSC, Victoria BC V8W 3P6, Canada\\
$^{6}$ Universit\"ats-Sternwarte M\"unchen, Scheinerstra{\ss}e 1, D-81679 M\"unchen, Germany\\
$^{7}$ Institute of Physics, Laboratoire d'astrophysique, \'Ecole Polytechnique F\'ed\'erale de Lausanne (EPFL), Observatoire, CH-1290 Versoix, Switzerland\\
$^{8}$ GEPI, Observatoire de Paris, Universit\'e PSL, CNRS, Place Jules Janssen, F-92190 Meudon, France\\
$^{9}$ New York University Abu Dhabi, PO Box 129188, Saadiyat Island, Abu Dhabi, United Arab Emirates\\
$^{10}$ Center for Astro, Particle and Planetary Physics (CAP$^3$), New York University Abu Dhabi
}

\date{Accepted XXX. Received YYY; in original form ZZZ}

\pubyear{2020}

\begin{document}
\label{firstpage}
\pagerange{\pageref{firstpage}--\pageref{lastpage}}
\maketitle

\begin{abstract}
The kinematics of the most metal-poor stars provide a window into the early formation and accretion history of the Milky Way. Here, we use 5~high-resolution cosmological zoom-in simulations ($\sim~5\times10^6$ star particles) of Milky Way-like galaxies taken from the NIHAO-UHD project, to investigate the origin of low-metallicity stars ($\FeH\leq-2.5$). The simulations show a prominent population of low-metallicity stars confined to the disk plane, as recently discovered in the Milky Way. The ubiquity of this finding suggests that the Milky Way is not unique in this respect. Independently of the accretion history, we find that $\gtrsim~90$ per cent of the retrograde stars in this population are brought in during the initial build-up of the galaxies during the first few Gyrs after the Big Bang. Our results therefore highlight the great potential of the retrograde population as a tracer of the early build-up of the Milky Way. The prograde planar population, on the other hand, is accreted during the later assembly phase and samples the full galactic accretion history. In case of a quiet accretion history, this prograde population is mainly brought in during the first half of cosmic evolution ($t\lesssim7$~Gyr), while, in the case of an on-going active accretion history, later mergers on prograde orbits are also able to contribute to this population. Finally, we note that the Milky Way shows a rather large population of eccentric, very metal-poor planar stars. This is a feature not seen in most of our simulations, with the exception of one simulation with an exceptionally active early building phase.
\end{abstract}


\section{Introduction}
The most chemically pristine stars, which likely include some of the oldest stars in the Milky Way, are relics of the early formation and assembly of our Galaxy \citep[e.g.,][]{Freeman02,Karlsson13}. For example, \citet{ElBadry18} show that, in the cosmological FIRE simulations, stars with $\FeH\leq-2.5$ formed at most 3 $\Gyr$ after the Big Bang. During that time, the Milky Way (hereafter MW) was still assembling. The expectation is therefore that these stars are distributed in  a pressure-supported fashion, \ie the spheroid; either located in the deepest part of the MW potential well if accreted at early times, or spread out to the outer reaches of the stellar halo if born in accreted dwarf galaxies \citep{White00, Brook07, Gao10, Salvadori10, Tumlinson10, Ishiyama16, Starkenburg17b, ElBadry18, Griffen18}. Some orbits from the pressure-supported distribution might be kinematically coincident with the disk, although for this distribution the number of prograde and retrograde stars would be expected to be similar. Alternatively, some of the low-metallicity stars can cross the disk with halo kinematics.

However, thanks to the exquisite  data from the  Gaia mission \citep[hereafter Gaia DR2]{Gaia16,Gaia18}, \citet[][hereafter S19]{Sestito19}  found that a surprisingly large fraction (11 stars, $\sim 26$ per cent) of the 42 ultra metal-poor stars (UMP,  $\FeH \leq -4.0$) known at the time do not venture far out from the MW plane ($|z_{\rm max}|\lesssim3\kpc$) and have orbits that span a wide range of eccentricities, from quasi-circular to rosette-shaped orbits. Out of the 11 stars confined to the disk, 10 UMPs have prograde motion, sharing the same sense of rotation as the MW disk, while one UMP has a retrograde orbit. \citet[][hereafter S20]{Sestito20} extended the kinematical analysis to  the very metal-poor regime (VMP, $\FeH\leq-2.0$) using 583 VMP stars from the Pristine survey \citep{Starkenburg17a,PristineDavid} and 4838 VMPs from a cleaned sample of the LAMOST survey \citep{Cui12,Li18}. S20 found that a similarly large fraction of the sample kinematically inhabits the plane of the MW, from the VMP to the UMP regime. They also show that the prograde motion is largely favoured  compared to  retrograde orbits. S19 and S20 propose three non mutually-exclusive scenarios to explain the observations. The first scenario is that this population was brought in by accretion events where satellites deposited their stars into the disk by dynamical friction and tidal interactions. In the second scenario, these stars were born in the gas-rich building blocks that formed the backbone of the proto-MW disk. Finally, in the last scenario, these stars formed in situ after the interstellar medium (hereafter ISM) of the disk settled, presumably from still chemically pristine pockets of gas.

\citet{DiMatteo20} analysed a sample of 54 VMP from the ESO Large Program "First Stars" finding very similar kinematical signatures to S19 and S20. They suggest that the MW thick disk extends to the UMP regime, down to $\FeH\sim-6.0$ and that the population of this early disk shares the same kinematical properties because it experienced the same violent heating process, \ie the accretion of the Gaia-Enceladus satellite \citep{Belokurov18,Helmi18}. \citet{Venn19}, analysing the high-resolution spectra of 28  bright low-metallicity stars ($\FeH\leq-2.5$), also found that a subsample is confined to the MW disk, sharing a wide range of eccentricities, with the majority in prograde motion.

Because it is clear that the retrograde stars cannot be easily explained in in-situ star formation scenarios, their presence --- even if they represent a minority population --- places important constraints on the early formation history of the galaxy. We note that, precisely for this reason, some of the important big merger events of the MW have been picked up from their retrograde signatures, such as, \eg Gaia-Enceladus-Sausage \citep{Belokurov18,Helmi18} and Sequoia \citep{Myeong19, Barba19} and Thamnos \citep{Koppelman19}. Additional information on the kinematical components of the MW seems to be encoded in the chemical abundances of disk stars \citep[e.g.][]{Navarro2011}. In fact, already prior to the Gaia mission such chemical peculiarities in combination with distinct kinematical features were used to identify debris stars associated with the merger event that brought $\omega$Cen into the MW, which is most likely coincident with Gaia-Enceladus-Sausage \citep[][]{Meza2005,Navarro2011}. \citet{Helmi18} and \citet{Koppelman19} demonstrate a different $\alfe$ between some accreted halo populations and the presumed in-situ population at a metallicity between $\sim-2.5$ and  $\sim-0.5$. \citet{Monty20} find that the stars dynamically associated  with Gaia-Enceladus-Sausage and Sequoia have a different  $\alfe$ ratio than those of the Galactic halo. For the most metal-poor stars, which are the subject of this work, the measurement of $\alfe$ and other elements tracers of the star formation (\ie neutron-capture elements) becomes challenging. Moreover, we know from previous works that differences between systems in $\alfe$ become less pronounced at very low metallicity \citep[e.g.,][]{Venn04,Jablonka15,FrebelNorris15}, while the chemistry of neutron-capture elements, such as Yttrium and Europium, has shown to be a promising accretion diagnostic \citep{RecioBlanco20}.

In this paper, we use the MW-like simulated galaxies present in the NIHAO-UHD\footnote{UHD stands for Ultra High Definition.} cosmological zoom-in simulations \citep{Buck20} to investigate how the oldest and most metal-poor stars assemble into the main galaxy. In particular, we focus on those stars that are confined to the disk at the present day. This suite is composed of 5 high-resolution spiral galaxies with $\sim200\pc$ resolution and are, therefore, ideal to disentangle the different structures of the galaxies \citep{Buck18,Buck19b}. The resolution allows us to analyse the spatial and kinematical distribution of the most metal-poor stars, focusing on the inner region of the galaxies. 

Section~\ref{simulations} describes the main properties of the NIHAO-UHD simulations, while, in Section~\ref{results}, the analysis and discussions on the origin of the most metal-poor stars are reported. In Section~\ref{spheroidsec}, we investigate the rotation of the VMP spheroids in the simulations. The comparison between the observations and the simulated galaxies is shown in Section~\ref{mwsim}. Section~\ref{growthsection} describes the formation of the simulated galaxy and their accretion history. The age-metallicity relation is shown in Section~\ref{agemetsection}, and the origin of the most metal-poor star particles is investigated in Section~\ref{trackingsection}. The conclusions drawn from this analysis are presented in Section~\ref{conclusions}.

\section{NIHAO-UHD Cosmological Zoom-in Simulations}\label{simulations}
The NIHAO-UHD simulations \citep{Buck20} are a set of cosmological simulations with a higher mass resolution and the same initial conditions and feedback parameters as the Numerical Investigation of a Hundred Astronomical Objects simulation suite \citep[NIHAO]{Wang15}. All the NIHAO galaxies adopt cosmological parameters from the \citet{Planck14} cosmology. Therefore $\Omega_{m} =0.3175$, $\Omega_{\Lambda}= 0.6825$, $\Omega_{b}= 0.0049$, $\mathrm{H}_0 = 67.1 \kms \Mpc^{-1}$, and $\sigma_{8} = 0.8344$. The corresponding age of the Universe is $t_{\rm Universe}=13.83\Gyr$. Each simulation consists of 64 snapshots equally spaced in time with a separation of $\sim215$ Myr. The final snapshots of the NIHAO-UHD simulations at $z=0$ are publicly available at \url{https://tinyurl.com/nihao-uhd} which redirects to \url{https://www2.mpia-hd.mpg.de/\~buck/\#sim_data}.

The NIHAO-UHD set is composed of six zoom-in simulations \texttt{g2.79e12}, \texttt{g1.12e12}, \texttt{g8.26e11}, \texttt{g7.55e11}, \texttt{g7.08e11} and \texttt{g6.96e11}, for which the name corresponds to the halo mass of the dark matter (hereafter DM) only run. The stellar mass of the galaxies varies between $1.5\times10^{10} \msun$ to $15.9\times 10^{10} \msun$. Each galaxy is resolved with more than $10^7$ particles (gas+star+DM) inside the virial radius while the stellar disks contain $\gtrsim3\times10^6$ star particles. The mass of DM particles spans a range between 1--5$\times 10^5 \msun$, gas particles between 2--9$\times 10^4 \msun$, and star particles between 0.7--3.0$\times 10^4 \msun$. The simulations have been evolved with a modified version of the smoothed particle hydrodynamics (SPH) code \texttt{GASOLINE2} \citep{Wadsley17} with star formation and feedback prescriptions as presented in \citet{Stinson2006} and \citet{Stinson2013}. The adopted feedback prescriptions result in a spatial distribution of young stellar particles in good agreement with the spatial distribution of young stellar clusters in local galaxies \citep{Buck19c}. Chemical enrichment from core-collapse supernova (SNII) and supernova Ia (SNIa) is implemented following \citet{Raiteri1996} using rescaled SNII yields from \citet{Woosley1995} and SNIa yields from \citet{Thielemann1986}. No Population III pre-enrichment is assumed and iron and oxygen have been tracked individually.

\begin{table*}
\caption{Properties of the simulated galaxies in NIHAO-UHD \citep{Buck20}. For each simulated galaxy, we report the masses of the single stellar, gas, and dark matter particles ($m_{\mathrm{star}}$, $m_{\mathrm{gas}}$, $m_{\mathrm{DM}}$), together with the total mass components (stellar $M_{\mathrm{star}}$, gas $M_{\mathrm{gas}}$,  and dark matter $M_{\mathrm{DM}}$), the total mass at the virial radius $M_{\mathrm{virial}}$, the virial radius $R_{\rm virial}$, the galaxy's disk scale length $R_d$, its thick disk scale height $h_{\mathrm{z, thick}}$ at the solar circle as defined via double exponential fit to the vertical stellar density, the mass of the VMP star particles within $40\kpc$ from the galactic centre, and its percentage relative to the total stellar mass.}
\label{nihaotab}
\resizebox{\textwidth}{!}{
 \begin{tabular}{c|c|c|c|c|c|c|c|c|c|c|c|c|}

\hline 
 Galaxy & $m_{\mathrm{star}}$ & $M_{\mathrm{star}}$ &   $m_{\mathrm{gas}}$ & $M_{\mathrm{gas}}$  & $m_{\mathrm{DM}}$ & $M_{\mathrm{DM}}$  & $M_{\rm virial}$  & $R_{\rm virial}$ & $R_d$  & $h_{\mathrm{z, thick}}$ &$M_{\rm VMP}^{40\kpc}$ & P(VMP) \\ 
 & $(10^4\msun)$ & $(10^{10}\msun)$ & $(10^4\msun)$ & $(10^{10}\msun)$ & $(10^5\msun)$ & $(10^{11}\msun)$ &  $(10^{12}\msun)$  &    $(\kpc)$ & $(\kpc)$ & $(\kpc)$ & $(10^{8}\msun)$ & (per cent)\\ \hline
\texttt{g2.79e12} & 3.13&  15.9& 9.38 & 18.48& 5.141& 27.90 & 3.13&  306 &5.57& 1.3& 12.25 &  0.77\\ 
\texttt{g8.26e11} & 1.32  & 3.40 & 3.96& 6.09 & 2.168 &8.26  &0.91 &206 &5.12 &1.4 & 1.39 &0.41 \\ 
\texttt{g7.55e11} &  0.93& 2.72& 2.78& 6.79& 1.523 &7.55  & 0.85&201 &4.41 &1.4 & 4.13 &1.52 \\ 
\texttt{g7.08e11} &0.68  & 2.00& 2.03& 3.74& 1.110 & 7.08 &0.55 & 174& 3.90& 1.0& 4.19 & 2.10  \\ 
\texttt{g6.96e11} & 0.93 &1.58 & 0.93& 4.79 & 1.523 & 6.96 & 0.68& 187 & 5.70 &1.4 & 3.47 & 2.20  \\  \hline
\end{tabular}}
\end{table*}

The NIHAO-UHD galaxies are characterised by a thin disk of scale length $\sim 5 \kpc$ and a total scale height of $\lesssim 1 \kpc$ that matches key observational properties of the MW, such as the age-velocity dispersion relation of the stellar disk \citep{Buck20} or the chemical bimodality of disk stars \citep{Buck20b}. Furthermore, one of the simulated galaxies, \texttt{g2.79e12}, has recently been used to study intrinsic variations in the length of the galactic stellar bar \citep{Hilmi20}. The galaxy \texttt{g1.12e12} has a spheroidal shape and is therefore not considered in this study. Table~\ref{nihaotab} reports the main properties of the NIHAO-UHD simulated galaxies used in our study. For a more detailed discussion of the galaxy properties we refer the reader to \citet{Buck19} and \citet{Buck20}.
 
For the simulated galaxy \texttt{g8.26e11}, several early snapshots have not been saved, complicating the tracking of the position of the star particles across time. Consequently, we exclude this galaxy when we are analysing snapshots at other times than the present day.

\section{Results and Discussions}\label{results}
The NIHAO-UHD simulations provide a set of physical quantities for each particle, such as their position at redshift 0, $(x,y,z)$, their Galactocentric velocity, $(v_x,v_y,v_z)$, their age, their metallicity, \FeH, and their birth position, $(x_{\rm birth},y_{\rm birth},z_{\rm birth})$. In all our analysis, we align the $z$-axis of the coordinate system with the total angular momentum of the disk stars such that the galactic disk lies within the $x-y$ plane.
Using the \texttt{AGAMA} package \citep{agama} we further derive the orbital action momentum vector, $(J_{\phi},J_{r}, J_{z})$ for each star particle corresponding to its orbit in the fixed gravitational potential of the simulated galaxy at redshift $z=0$.
 
In order to compare the population of planar stars in the simulations with that of the MW observations, as in Section~\ref{mwsim}, we need to mimic the window function of the photometric and spectroscopic surveys used to discover them. A  deep analysis and reconstruction of the selection functions is not the aim of this work; however, we mimic the window function of the observed VMP stars in S20, which also contain the stars analysed in S19, by selecting metal-poor star particles that are close to the location of stars in the observed sample. As also discussed in S19 and S20, the multiple selection functions imparted by different metal-poor surveys do not insert a bias against or in favour of the prograde/retrograde population. Moreover, mimicking the spatial window function is not expected to result in mimicking the kinematical features of the MW. This is also not desired, as we want to study these properties and not set them a priori. Because of the arbitrary choice of the orientation of the Galactocentric cartesian axes in the simulations, for each observed VMP star at position $(x_{\rm obs},y_{\rm obs},z_{\rm obs})$, we first select all the VMP star particles that inhabit a torus with $R_{\rm torus} = (x_{\rm obs}^2 + y_{\rm obs}^2)^{1/2}$, we set the width of the torus to $0.75 \kpc$, at the height $z=z_{\rm obs}$.  Usually, multiple star particles populate this volume, in which case the final choice of particle is done selecting the one inside the torus that minimises the difference in metallicity $|\FeH_{\rm obs}-\FeH_{\rm particle}|$. Since the observed samples in S19 and S20 exclude the bulge and the bar of the MW (see Appendix in S20), the orientation of the bar in the simulated galaxies does not influence this attempt in reproducing  the window function. The mimicking of the window function only applies to Section~\ref{mwsim}, where we compare more directly with observations, while for Sections~\ref{growthsection},~\ref{agemetsection},~and~\ref{trackingsection} all VMP stars ($\FeH\leq-2.0$) that are taken into account have a distance from the centre of the galaxy $R\leq 40 \kpc$ and exclude the bulge region ($R\geq 4 \kpc$) . 

\subsection{Do the low-metallicity stars confined to the disk follow a spheroidal distribution?}\label{spheroidsec}
The most metal-poor stars mainly inhabit the spheroid of the MW. However, as recently pointed out by observations \citep[\eg][]{Sestito19,Sestito20,DiMatteo20}, a non-negligible fraction of these stars is kinematically confined to the disk, favouring the prograde motion. Naturally, even if all stars were to be distributed in a non-rotating or a slowly-rotating spheroid, one would expect a subset of them to overlap with the disk at any time, and an even smaller subset to be confined to the disk kinematically. Benefitting from the completeness of our simulations --- in which, in contrast to observations, we can assess the complete metal-poor population --- it is insightful to quantify if the number of metal-poor stars kinematically confined to the disk exceeds the expectations from a spheroidal (rotating or non-rotating) distribution. 

\begin{figure*}
\includegraphics[width=\hsize]{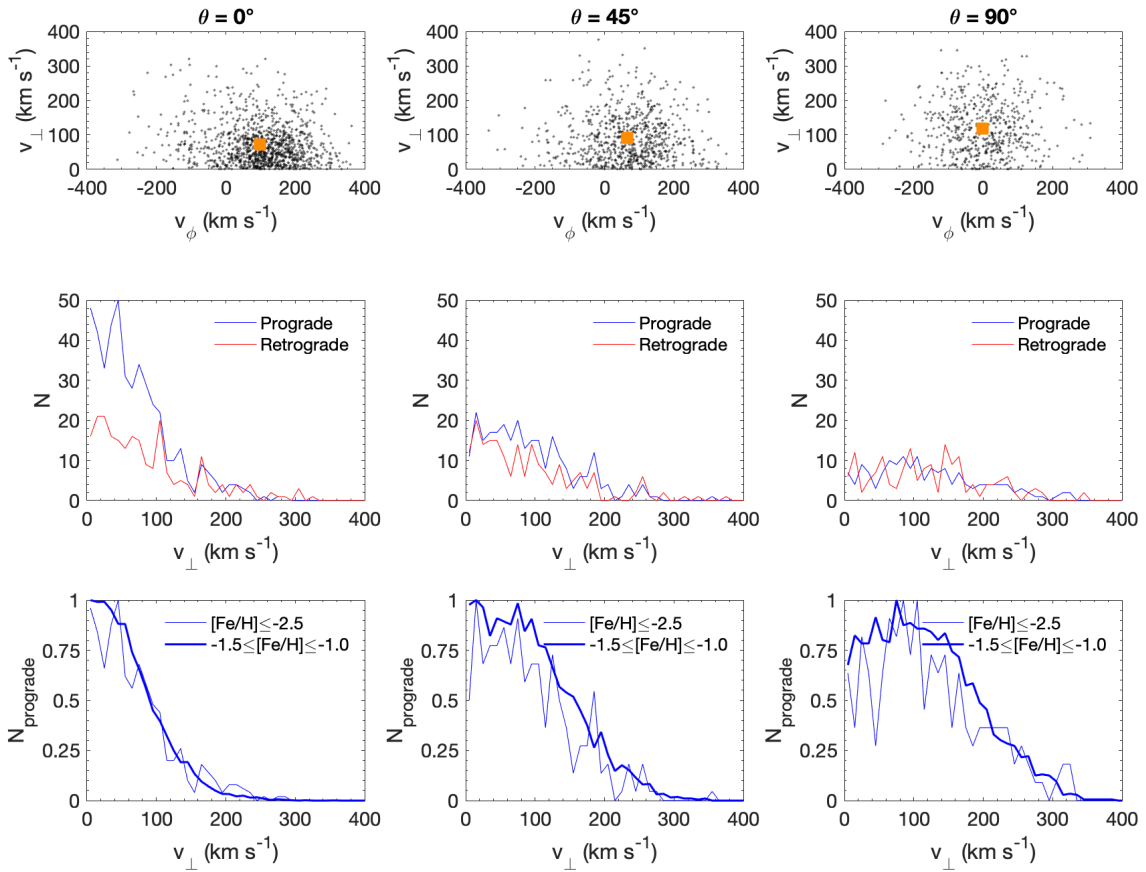}
\caption{Test of a spheroidal distribution for simulation \texttt{g8.26e11}. Top panels: rotational velocity $v_{\phi}$ vs. perpendicular velocity $v_{\perp} = |v_z|$ for the star particles located in the disk ($|z|\leq 3\kpc,R\geq4\kpc$) for three rotational angle ($0,45,90\degree$) and with $\FeH\leq-2.5$. The dark orange marker represents the centroid of the distribution in this space. Middle row panels: the number of prograde and retrograde stars as a function of the perpendicular velocity is shown in blue and red, respectively, and for the three rotational angles ($0,45,90\degree$). These distributions have been corrected by the mean rotation of the spheroid and selecting star particles with $50\leq v_{\phi}\leq250\kms$ for the prograde and $-250\leq v_{\phi}\leq-50\kms$ for the retrograde. Bottom panels: the comparison between the number of the low-metallicity prograde ($\FeH\leq-2.5$, thinner line) and the more metal-rich prograde population ($-1.5\leq\FeH\leq-1.0$, thicker line) as a function of $v_{\perp}$.}\label{spheroid}
\end{figure*}

To this end, we perform the following exercise: we select the low-metallicity star particles ($\FeH\leq-2.5$) of the NIHAO-UHD simulations that are spatially in the stellar disk $|z|\leq3 \kpc$ (\ie close to the $x-y$ plane of the coordinate system), excluding the bulge region ($R\geq4\kpc$). Their distribution in the rotational velocity $v_{\phi}$ vs. the perpendicular velocity $v_{\perp} = |v_{z}|$ space is shown in the top-left panel of Figure~\ref{spheroid} and indicates that the spheroid has a slow prograde rotation in this simulation (\texttt{g8.26e11}). This signal is consistently present in the 5 simulations studied in this analysis. In the top-middle and top-right panels of Figure~\ref{spheroid}, the coordinate system is rotated around the x-axis, such that the stellar disk moves out of the $x-y$ plane and onto an angle of $45\degree$ and $90\degree$ respectively. In this new frame of reference, we apply the same selection of star particles. As the rotational angle increase, the mean rotational velocity of the spheroid decreases, illustrating that the rotation is indeed strongest in the $x-y$ plane. This rotation will lead to star particles being dragged towards the disk and co-rotate with it, although with smaller rotational velocity than the disk \citep[see Figure~7 from][for the rotation curve of these simulated galaxies]{Buck20}.  

In the second row of panels in Figure~\ref{spheroid}, we correct the rotational velocity by the mean rotation of the population and we measure the number of prograde and retrograde star particles as a function of the perpendicular velocity. These two populations are selected on their corrected rotational velocity, in the range $50\leq v_{\phi}\leq250 \kms$ for the prograde star particles and in the same range with opposite signs for the retrograde star particles. In case of a spheroidal distribution around the mean rotation, the fraction of prograde and retrograde orbits should be equal and constant regardless of whether the chosen plane coincides with the galactic disk or is at an angle. We however see that at $0\degree$, there is a predominance of prograde metal-poor stars confined to small perpendicular velocities (i.e., not venturing far from the disk plane). For planes rotated out of the disk (second and third panels in the middle row of Figure~\ref{spheroid}), the number of retrograde and prograde stars is similar and more weakly depends on the perpendicular velocity. 

The bottom panels in Figure~\ref{spheroid} show a comparison between the low-metallicity prograde population and the more metal-rich prograde population that contains the thick disk ($-1.5\leq\FeH\leq-1.0$). These panels indicate that the low-metallicity population has a similar distribution to that of the more metal-rich star particles, especially for low perpendicular velocities and for small rotational angle. This means that the population of low-metallicity prograde planar star particles behaves similarly to the thick disk, and likely exceeds the expected number of star particles drawn from a simple, slowly-rotating spheroidal distribution. 

The preference for prograde motion among the very metal-poor stars in the region of the simulated disk is qualitatively in agreement with the observations \citep[\eg][]{Sestito19,Sestito20}. While the simulations host spheroids that are clearly rotating, the mean motion of the MW is still under debate. Recently, thanks to the LAMOST survey \citep{Cui12}, \citet{Tian19,Tian20} found that the MW spheroid is slowly-rotating in a prograde motion ($v_{\phi}\sim 29 \kms$) and that the rotational velocity decreases at larger distances. Accreted structures in the MW halo, either retrograde and prograde, complicate the estimation of the rotational velocity of the spheroid and can lead to different results depending on the tracers used \citep{Deason11}.

In conclusion we see that, despite the more significant rotation of the spheroid in these simulations compared to the MW, there is a clear signature on top of that of a very low-metallicity population that spatially and kinematically reside in a (thicker) disk, just as we see in the MW galaxy. Although none of these simulations exactly resemble the MW in its formation history, there is still a lot we can learn from studying the origin of this population that seems to be ubiquitous throughout the different simulated galaxies. In the next section, we will provide a more in-depth comparison between the types of orbits these star particles have in the simulations and the observations in the MW and we will pursue a deeper understanding of the origin of these populations with different orbital properties in Section~\ref{trackingsection}. 

\subsection{NIHAO-UHD simulations vs. the observed low-metallicity MW disk}\label{mwsim}
The action-angle variables (hereafter action vector, or action) are a useful tool to analyse dynamical populations. In particular, the action coordinates $J_{\phi}$ and $J_{z}$ can reveal a population of the VMP stars in the disk of the MW, as highlighted by  S19 and S20.

\begin{figure*}
\includegraphics[width=\hsize]{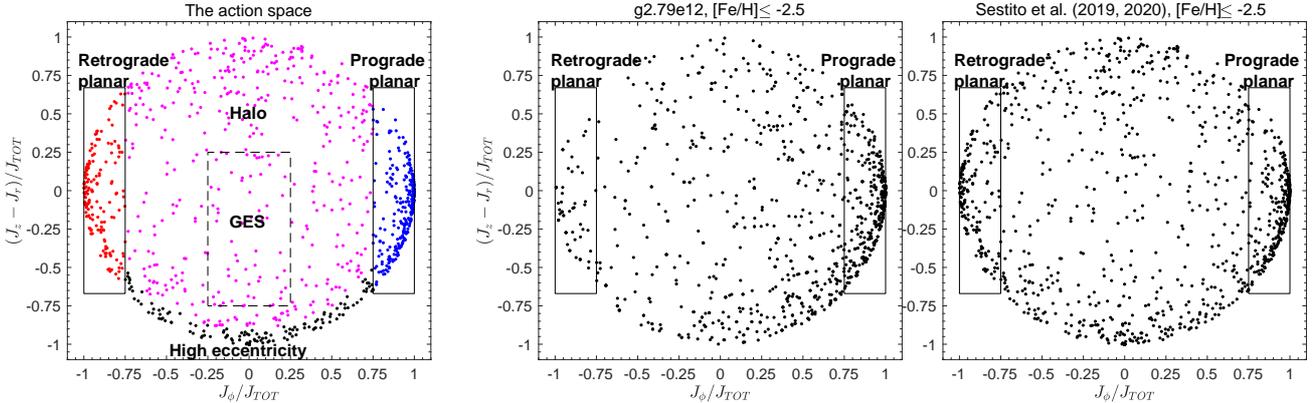}
\caption{Action momentum space for star particles in the simulation and stars in the MW. Left panel: Sketch of the action space. The $x$-axis shows the azimuthal component of the action vector $J_{\phi}/J_{\rm TOT}$, a prompt for the rotational motion. Prograde orbits have $J_{\phi}/J_{\rm TOT}\geq 0$, while star particles with retrograde  motion have $J_{\phi}/J_{\rm TOT}<0$. The $y$-axis, $(J_{z}-J_r)/J_{\rm TOT}$, is the difference between the vertical component, which tracks the vertical motion of the particle/star, and the radial component of the action vector, which is an indication of its radial motion. Both axes are normalised by the norm of the action vector, $J_{\rm TOT}$. This is helpful for a comparison between galaxies with different physical properties. The black boxes represent the loci we define for prograde planar (right box, also shown with blue dots) and the retrograde planar stars (left box, also marked by red dots). The halo-like star particles are denoted by magenta dots, while the star particles with high eccentricity are marked by black dots. The dashed-line box represents the region where Gaia-Enceladus-Sausage (\ie GES) has been discovered \citep{Belokurov18,Helmi18}.
Central panel: the star particles with $\FeH\leq-2.5$ from simulation \texttt{g2.79e12} selected to mimic observations. Right panel: MW observations from S19 and S20. In the both the central and right panel, the star particles are marked with black dots.}\label{action1}
\end{figure*}

The simulated galaxies are not models of the MW. This means that the kinematic of star particles may differ systematically from the MW's because of differences in the mass and spatial distribution of the DM, stars, and gas. To minimise these effects, we scale the components of the actions by their norm. This is illustrated in Figure~\ref{action1}, where we compare the action space of \texttt{g2.79e12} (central panel) with that of the MW observations from S19 and S20 (right panel). The VMP star particles in the simulation are selected in the aforementioned way to mimic the observational window function. This plot shows the azimuthal action component  $J_{\phi}/J_{\rm TOT}$ versus the difference between the vertical and the radial action component $(J_{z}-J_r)/J_{\rm TOT}$, where both axes are normalised by the norm of the action vector $J_{\rm TOT}$. In this action space, stars with planar orbits, prograde and retrograde, inhabit regions with high $|J_{\phi}/J_{\rm TOT}|$ and low $|(J_{z}-J_{r})/J_{\rm TOT}|$, repsectively. Therefore, we define the star particles with prograde and planar orbits to be confined in the region with $J_{\phi}/J_{\rm TOT}\geq0.75$ and the star particles with retrograde motion that are confined in the disk to have $J_{\phi}/J_{\rm TOT}\leq-0.75$. At the bottom of this space are found star particles that are confined to the disk but have very radial orbits (\ie eccentric) orbits. Star particles with halo-like kinematics inhabit the remainder of the space.
  
  \begin{figure*}
\includegraphics[width=\hsize]{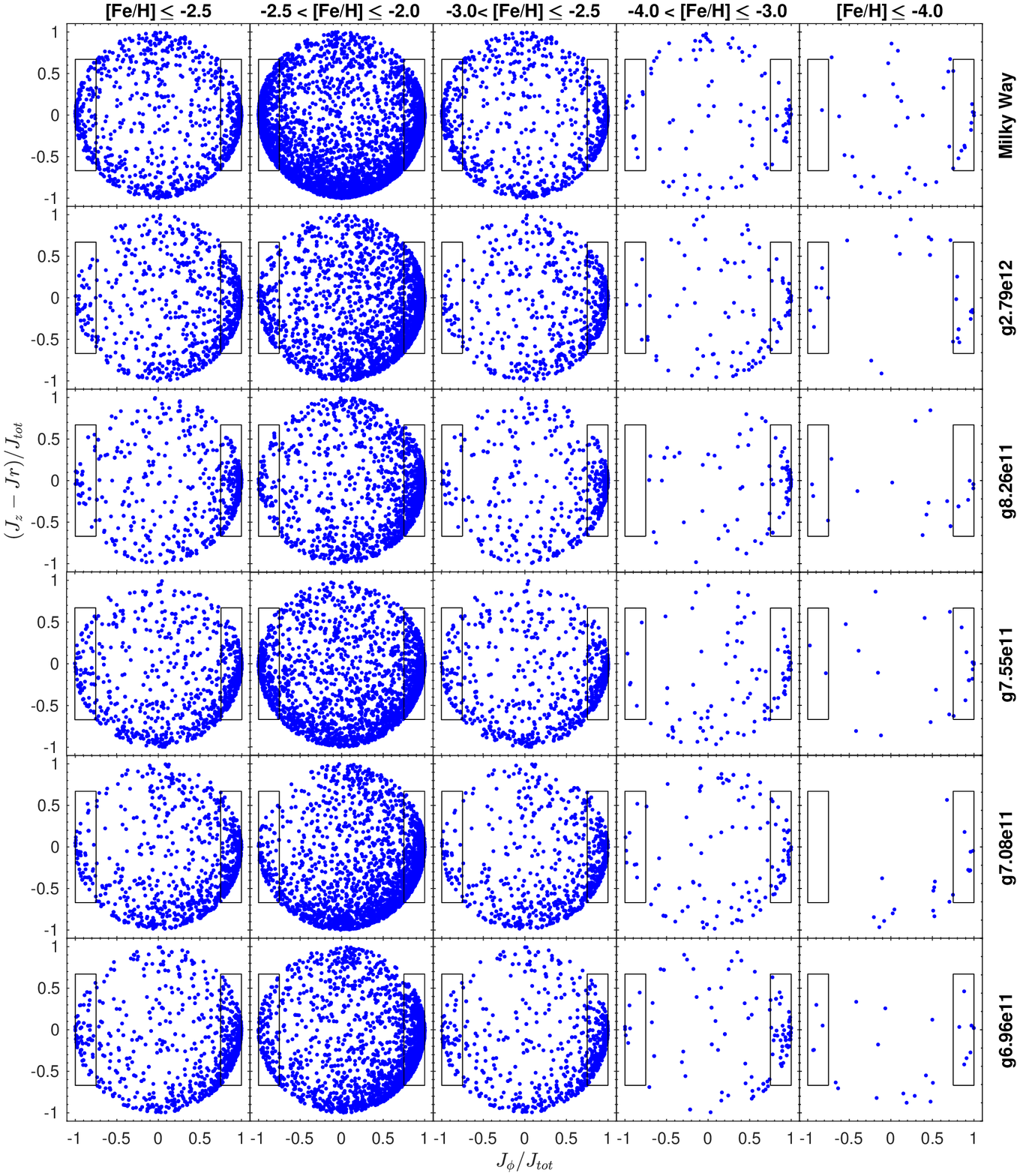}
\caption{Metallicity decomposition of the simulated galaxies and MW observations in action space. The first row of panels displays the  observations from S19 and S20, while the other rows show the simulated galaxies of the NIHAO-UHD suite, corrected for the observational window function as explained in the text. Each row is divided into metallicity bins. To better visualise the  population of the most metal-poor star particles ($\FeH\leq-2.5$), the left column of panels shows all stars in the most metal-poor sample. From the second to the fourth column, the sample of the VMP stars has been separated according to their metallicity, from the VMP to the UMP regime. In each panels, the black boxes on the right and on the left represent the loci the populations of prograde planar and retrograde planar star particles.}\label{action2}
\end{figure*}

\begin{table*}
\caption{Prograde vs. retrograde planar asymmetry. The ratio $N_{\rm pro}/N_{\rm retro}$ between the number of star particles with prograde/retrograde planar orbits is reported as a function of the metallicity range for the simulated galaxies and the observed VMP stars in the MW \citep{Sestito20}.}
\label{ratio_table}
\resizebox{\textwidth}{!}{
\begin{tabular}{ccccccc}
\hline 
Galaxy& $\FeH\leq-2.5$            & $ -2.5\leq\FeH\leq-2.0$      & $ -3.0\leq\FeH\leq-2.5$   & $-4.0\leq\FeH\leq-3.0$  &   $\FeH\leq-4.0$  \\ \hline
Milky Way &  1.72 & 1.89 & 1.67 & 1.82 & 11.11    \\ 
\texttt{g2.79e12} &  7.14 & 9.09 & 8.33 & 6.25 & 2.33    \\ 
\texttt{g8.26e11} &  9.09 & 12.50 & 8.33 & -- & 2.33   \\ 
\texttt{g7.55e11} &  3.45 &3.13 & 3.23 & 6.25 & 7.14    \\ 
\texttt{g7.08e11} &  6.25 & 7.69 & 6.67 & 5.00 & --   \\ 
\texttt{g6.96e11} & 5.56 & 7.69 & 5.88 & 7.14 & 2.00    \\  \hline
\end{tabular}}
\end{table*}

Figure~\ref{action2} shows the same action space as in Figure~\ref{action1} but divided in metallicity bins for all the NIHAO-UHD simulated galaxies\footnote{For \texttt{g8.26e11}, the action-angle variables are safe to use as their calculation is independent of the snapshots at the early Universe.} as well as for the MW. From this plot, a population of planar stars is clearly recognisable in all simulations, from the VMP samples to the most metal-poor samples.  The action space of Figures~\ref{action1} and~\ref{action2} show that this population of planar star particles in prograde motion spans a wide range of eccentricities (from low to high values of $J_r$, inhabiting the lower border of the action space). Moreover, as with the MW observations \citep{Sestito19,Sestito20}, the simulated galaxies show some stars that are confined to the disk with retrograde orbits. Similarly to observations, the prograde sample is more populated than the retrograde one for all galaxies and at all metallicities. Table~\ref{ratio_table} reports the ratio between the prograde and retrograde planar populations, $N_{\rm pro}/N_{\rm retro}$, as a function of the metallicity and the simulated galaxy. In most simulated galaxies this ratio is $>5.50$, except for \texttt{g7.55e11}  which has a lower ratio of $\sim 3.4$. These numbers are significantly higher than what is observed for the MW ($\sim 1.7$), as also reported in Table~\ref{ratio_table}. This indicates that the simulations has an even larger population of prograde stars at low metallicity. As we already pointed out, each photometric and spectroscopic survey has its own selection function for hunting metal-poor stars; however, none of them should impart a bias for/against retrograde and prograde population. In Section~\ref{spheroidsec}, we report that the spheroids of the simulated galaxies are slowly rotating in a prograde motion while the MW spheroid is, at best, slowly rotating. This difference may impact the direct comparison of the $N_{\rm pro}/N_{\rm retro}$ ratios.

Careful comparison between the observed metal-poor MW and the NIHAO-UHD simulated galaxies in Figure~\ref{action2} reveals another interesting feature  in the lower hemicircle of the action space ($(J_z-J_r)/J_{\rm TOT} \lesssim 0$, $-1 \leq J_{\phi}/J_{\rm TOT}\leq1$), outside of the black boxes. Looking at the MW panels, there is a pronounced overdensity of stars in this area, a feature that is not matched in most of the simulations except (qualitatively) in \texttt{g7.55e11}. In this locus of action space we find stars that have large motion in the radial component (large $J_r$), compared to a smaller motion on the vertical axis ($J_z$), therefore these are planar stars with high eccentricity, the majority of them on prograde orbits. A more in-depth discussion on the origin of this orbital structure is presented below.

\subsection{Growth history of the galaxies}\label{growthsection}

\begin{figure}
\includegraphics[width=\hsize]{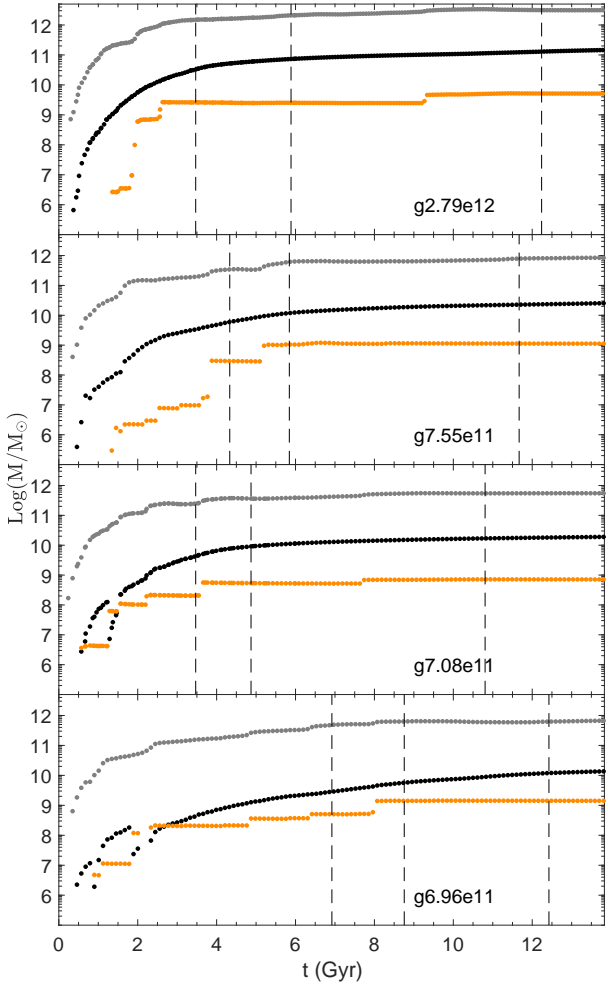}
\caption{Virial mass (gray), main halo stellar mass (black), and accreted stellar mass (dark orange) as a function of time. The stellar mass in a sphere with a radius of $r=75\kpc$ centred in the main halo is denoted by black dots, the total accreted stellar mass from satellites within the same radius is marked with dark orange dots, and the total virial mass of the main halo is shown by gray dots. Vertical lines represent the time at which the stellar component of the main halo reached $25$ per cent (left vertical line), $50$ per cent (middle vertical line), and $90$ per cent (right vertical line) of the total stellar mass. As previously mentioned, simulation \texttt{g8.26e11} is not shown as no reliable distinction between accreted and in-situ stars can be made. Note that the apparent reduction in main halo stellar mass for simulations \texttt{g7.08e11} and \texttt{g6.96e11} at very early cosmic times is due to the ongoing major mergers at those times.}\label{tree}
\end{figure}

Tracking the haloes and their properties such as the mass (DM, gas, and stellar components) and position is helpful to better understand how the simulated galaxies grew. The total stellar mass within the virial radius of the main halo, the total stellar mass of accreted material (also measured within the virial radius of the dwarf galaxies prior to accretion), and the virial mass as a function of the cosmic time are shown in Figure~\ref{tree}. For each simulated galaxy, the time at which the simulation assembles 25 per cent ($t_{25}$), 50 per cent ($t_{50}$), and 90 per cent ($t_{90}$) of their present stellar mass is indicated by vertical lines. Simulations \texttt{g2.79e12}, \texttt{g7.55e11}, and \texttt{g7.08e11} reach $t_{25}$ and $t_{50}$ after $\sim4\Gyr$ and $5-6\Gyr$, respectively. On the other hand, for simulation \texttt{g6.96e11}, this happens at $\sim 7\Gyr$ and $\sim 9\Gyr$. This can be explained by a more continuous accretion of mass in this latter simulation, for which the total accreted mass is a large fraction of the final galaxy \citep[see also Figure~3 of][]{Buck20}.

\begin{figure*}
\includegraphics[width=\hsize]{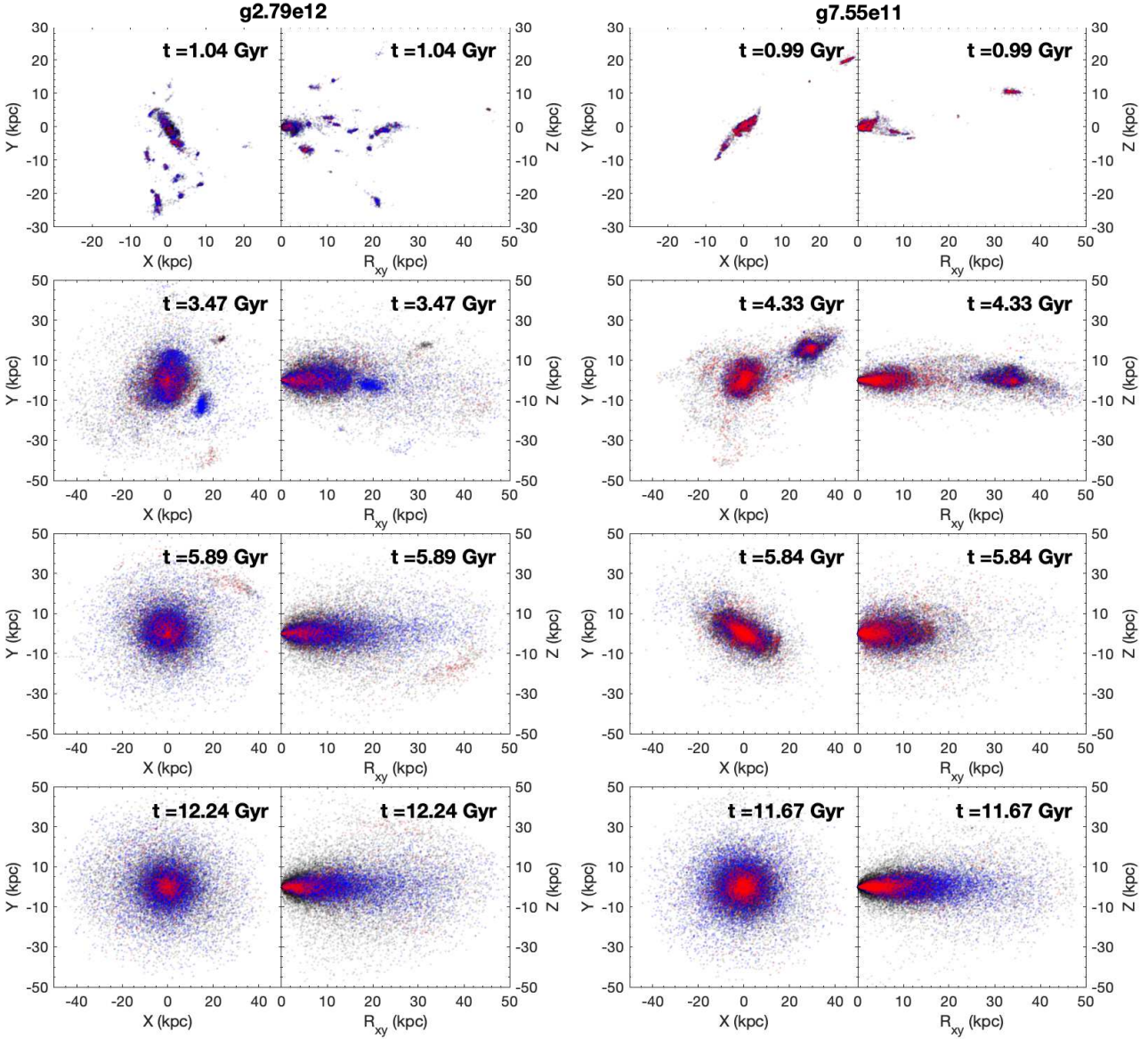}
\caption{Snapshots of the VMP particle distribution. Projection of the spatial distribution of the VMP star particles along the $x-y$ and $R_{\rm xy}-z$ planes for the simulations \texttt{g2.79e12} (on the left) and \texttt{g7.55e11} (on the right) at four different time ($t\sim1\Gyr$, $t_{25}$, $t_{50}$, $t_{90}$). Star particles with prograde planar orbits, retrograde planar motion, and halo-like distribution are denoted by blue, red, and black dots, respectively. To better visualise the component with the smaller number of particles, the halo-like star particles (black) have been plotted first, then the prograde planar components (blue), and the retrograde planar stars (red) have been overplotted on top of the others. In the right hand panels for \texttt{g7.55e11} at time $t_{25}$ ($t=4.33\Gyr$), one of the two massive mergers responsible for depositing star particles in the bottom part of the action space (see discussion in Sections~\ref{mwsim}~and~\ref{growthsection}) is visible at $x\sim30\kpc$, $y\sim20\kpc$, $z\sim0\kpc$, $R_{\rm xy}\sim35\kpc$.}\label{snap}
\end{figure*} 

Figure~\ref{snap} shows the projection on the $x-y$ plane and the $r-z$ plane (where the coordinate system is chosen such that the $z$-axis aligns with the total angular momentum of the stellar disk) of the VMP particle's position at 4 different time-frames for the simulated galaxies \texttt{g2.79e12} and \texttt{g7.08e11}. Figures~\ref{tree} and~\ref{snap} show that, in the first few $\Gyr$, the galactic building blocks (of which the most massive will be the main MW progenitor) of mass $10^5-10^9\msun$ merge together and assemble the proto-galaxy. These building blocks, bring in the most metal-poor star particles, together with the gas and the DM. Once the proto-galaxy is assembled, other merger events are responsible for bringing in more of the VMP star particles in the main structure. The number of late mergers and the mass they contribute varies from simulation to simulation, from a quiet accretion history after $\sim4$ Gyr for \texttt{g2.79e12} and \texttt{g7.08e11} to a more  turbulent and continued merging history for \texttt{g6.96e11} and \texttt{g7.55e11}.

As noted before,  simulation \texttt{g7.55e11} matches best the MW orbital property observations. This is true both for the ratio of metal-poor prograde vs. retrograde star particles, as well as for the existence of a significant population of high eccentricity planar star particles of low metallicity (see Figure~\ref{action2}). Simulation  \texttt{g7.55e11} stands out from the others in its very active and chaotic early merging phase in which more building blocks are coming together than in other simulations. Tracing the star particles belonging to this high eccentricity feature in \texttt{g7.55e11} back in time, we find that these belong to multiple merger events. Only $\sim20$ per cent of them are formed within $50\kpc$ of the center of the main halo, while the remaining particles were born in multiple satellites initially up to a distance of $300\kpc$. These later merge with the main galaxy. In particular, Figure~\ref{tree} shows that two massive satellites are merging at times  $\sim4\Gyr$ (also visible in Figure~\ref{snap}) and $\sim7\Gyr$. The stellar masses of the merging satellites are  $M_{\rm stellar} = 3.0\times10^8\msun$ ($M_{\rm TOT} = 6.7\times10^{10}\msun$) and $M_{\rm stellar} = 1.3\times10^9\msun$  ($M_{\rm TOT} = 6.7\times10^{10}\msun$), respectively. In both cases, the total mass (DM$+$gas$+$stars) of the merging satellites is about $\sim$20--25 percent of the main halo's mass. Before the first merger, $\sim24$ ($\sim55$) percent of the prograde (retrograde) star particles present in the high eccentricity feature are already in place as the result of this early merging phase. The first merger, accreted at the end of the building blocks phase, is responsible for bringing in $\sim22$ ($\sim40$) percent of the progrades (retrogrades) into the high eccentricity component, while the second merger has brought in  $\sim46$ per cent of the prograde and none of the retrograde star particles. Exploring the Auriga simulations, \citet{Gomez17} found that later massive merging events ($M_{\rm TOT} \sim 10^{10}$--$10^{11}\msun$) are able to accrete stars with kinematics that resemble the disk population. This is in line with the two later massive merging events present in simulation \texttt{g7.55e11}.

It is clear that the MW experienced an active merging episode in its history, as evidenced by the recent discoveries of mainly the Gaia-Enceladus-Sausage structure and several other structures (e.g., Gaia-Sequoia and Thamnos) that might or might not be related \citep[see, e.g., the review presented in][]{Helmi20}. Our finding that the significant population of high eccentricity planar star particles of low metallicity in the MW is not reproduced in all simulations, but rather might be the outcome of a more particular merger history, shows that this population in particular will be interesting to take into account when studying the merger picture of the MW in more detail. High-resolution spectroscopy of the low-metallicity planar stars, both prograde, retrograde, and highly eccentric, could provide further information on the  precise formation history of the MW.

\subsection{How old are the most metal-poor stars?}\label{agemetsection}
Since the abundance of metals in the ISM of the Galaxy increases gradually with successive generations of stars, the expectation is that VMP stars must have formed at early times, when the ISM was still relatively unpolluted. However, it is theoretically also possible that they form later from pockets of very isolated, unpolluted gas.

\begin{figure*}
\includegraphics[width=\hsize]{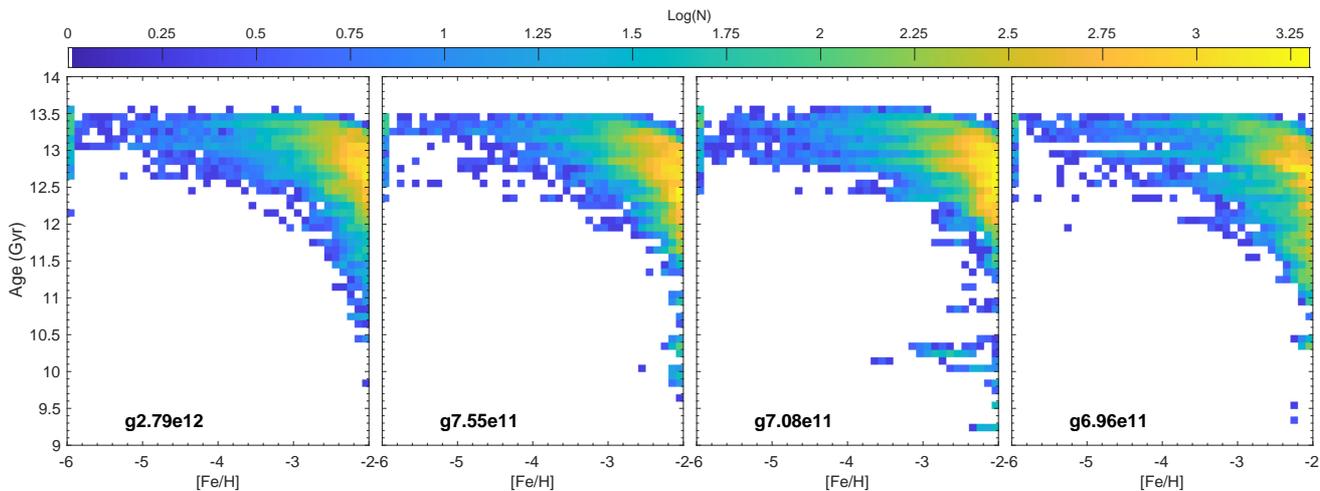}
\caption{Age vs. $\FeH$ for the VMP star particles in the simulated galaxies. Each pixel of size $0.1\Gyr\times0.1\rm{dex}$ is colour-coded by the logarithm of the particle counts. Star particles with $\FeH<-6$ are reported in the pixels of $\FeH=-6$ for a better representation. A younger population of low-metallicity stars is visible in the panel of simulation \texttt{g7.08e11}, as discussed in Section~\ref{agemetsection}, these star particles are brought in by a dwarf satellite, in which the older population has not been efficient in polluting the ISM, or pockets of pristine gas were accreted at a later moment. Tracking problems in simulation \texttt{g8.26e11} might have affected the age, and to be cautious, \texttt{g8.26e11} has been removed.}\label{metage}
\end{figure*}

Figure~\ref{metage} shows the distribution of the age of all the VMP star particles in the NIHAO-UHD simulations as a function of their metallicity. The metallicity-age plot is divided in a grid of $0.1$ dex in metallicity and $0.1\Gyr$  in age and colour-coded by the number density of star particles in each pixel. Overall, the simulations indicate that indeed the most metal-poor stars are also the oldest ones. Only 0.7--12.0 percent of all VMP star particles are younger than $12\Gyr$, while 35--77 percent are younger than $13\Gyr$. Their minimum age is, across all simulations, $\gtrsim 9.1\Gyr$. In the extremely metal-poor regime (EMP, $\FeH\leq-3.0$), 19--42 per cent of stars are younger than $\sim13\Gyr$, while only a few star particles are younger than $12\Gyr$ ($\leq 0.7$ per cent). Similarly, in the UMP regime, only 11--36 per cent are younger than $\sim13\Gyr$, and $\leq 0.7$ per cent are younger than $12\Gyr$. This result is in agreement with findings from the APOSTLE \citep{Starkenburg17b} and FIRE simulations \citep{ElBadry18} of MW-like galaxies.

While most of the VMP stars are thus truly old, Figure \ref{metage} also shows a population of stars in simulation \texttt{g7.08e11} with $\FeH<-2.5$ and an age $\leq10.5 \Gyr$. Selecting this younger subsample of star particles, we find that they were born in the same dwarf galaxy ($M_{\rm tot} = 4.89\cdot 10^{10}\msun$, $M_{\rm star} = 1.95\cdot 10^8\msun$, and $M_{\rm gas} = 5.46\cdot 10^9\msun$) that entered the virial radius ($R_{\rm virial}=174\kpc$ for  \texttt{g7.08e11}) at time $t\sim 5.84\Gyr$ after the Big Bang. This younger population of low-metallicity star particles is not born during the first peak of star formation in this small galaxy (there are much older stars present in this system), but the older population has not been efficient in polluting the ISM to higher metallicity. Alternatively, an infall of new, chemically pristine gas has occurred between the star formation episodes.

\subsection{Where do the most metal-poor stars come from?}\label{trackingsection}
An important goal of this paper is to answer the question of where the low-metallicity stars come from in order to distinguish between the three scenarios put forward in S19 and S20 (in short: later minor merging, came in with the early building block phase, or later in-situ formation in a quiescent disk). To do so, the NIHAO-UHD simulations provide the birth position $(x_{\rm birth},y_{\rm birth},z_{\rm birth})$ of the star particles, and, from each snapshot, we can track the position of the particles as a function of time $(x(t),y(t),z(t))$. With these quantities, it is possible to reconstruct the history of the most metal-poor stellar populations and connect the present kinematical properties to their counterparts at high-redshift.

From the previous analysis of the age-metallicity relation for the most metal-poor star particles (see Figure~\ref{metage}), it becomes clear that the population of star particles with $\FeH\leq-2.5$ is also the oldest ($\geq12\Gyr$), with almost no younger contaminants (see Section~\ref{agemetsection} for the discussion on the youngest population in \texttt{g7.08e11}). For this reason, we select this low-metallicity population ($\FeH\leq-2.5$) to better investigate when they were formed and accreted onto the main galaxy. We pay particular attention to the star particles that end up in prograde and retrograde planar orbits.

\begin{figure*}
\includegraphics[width=\hsize]{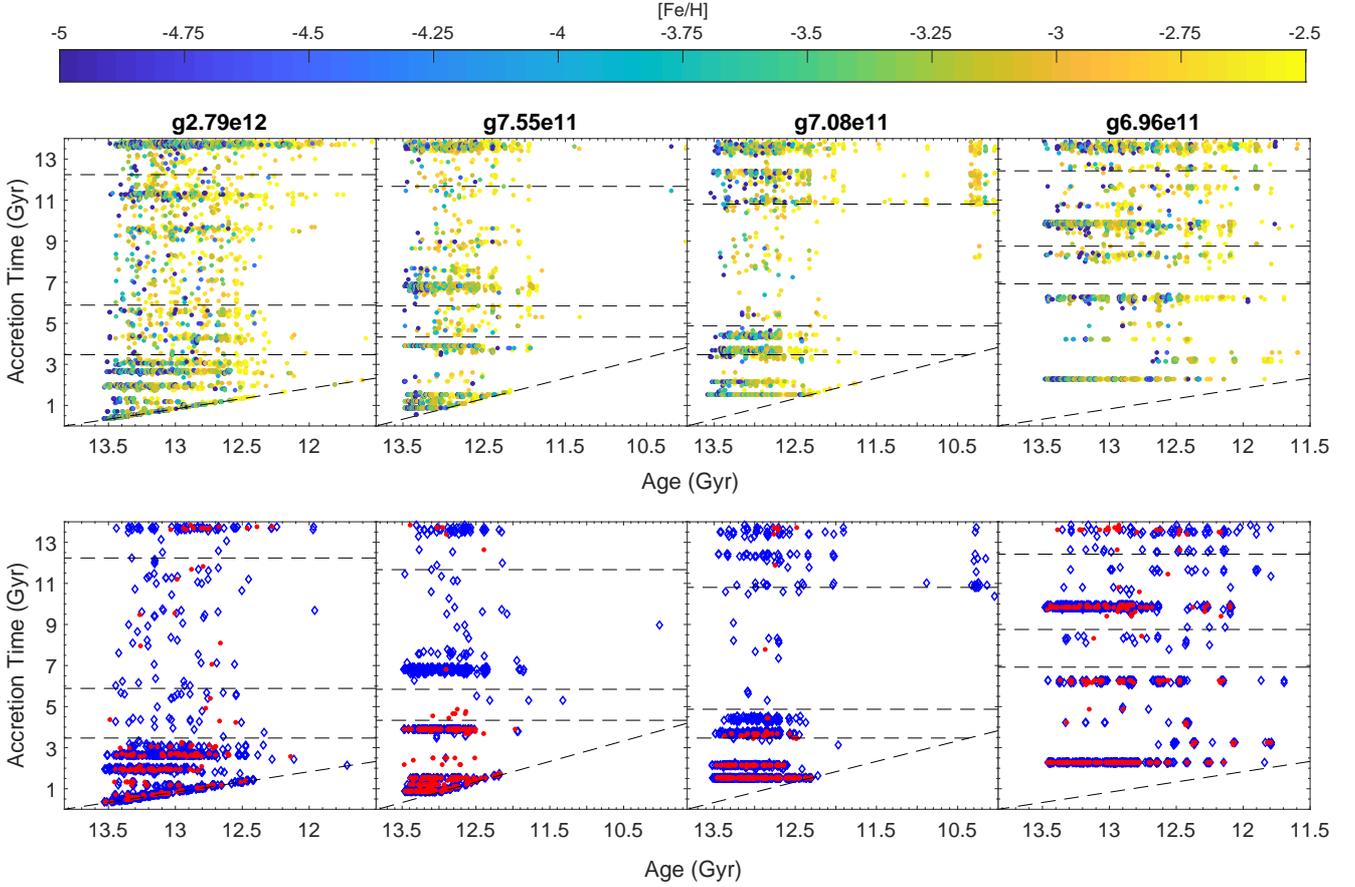}
\caption{Accretion times of the low-metallicity star particles. Top panels: Accretion time vs. age for low-metallicity ($\FeH\leq-2.5$) star particles within a Galactocentric radius of $50\kpc$ colour-coded by $\FeH$. Bottom panels: the accretion time vs. the age of the star particles as above, but now only for prograde planar star particles (displayed as blue rhombi) and retrograde planar particles (marked with a red dot). The horizontal dashed lines represent the time at which the main halo reached the $25$ per cent (lower line), the $50$ per cent (middle line), and the $90$ per cent (upper line) of the total stellar mass. The region below the inclined dashed line is forbidden since star particles would have been accreted before their formation. Star particles that lie on the inclined line have a birth position below $50\kpc$. Simulation \texttt{g8.26e11} has been removed due to tracking problems.}\label{injection}
\end{figure*}

Figure~\ref{injection} shows the time at which the low-metallicity star particles first enter a sphere of radius $50 \kpc$ centred on the main halo as a function of the age of the particle. In this figure, the over-densities of star particles with the same accretion time reflect the accretion of satellites. When these small galaxies merge with the MW-like galaxy, they  often deposit stars with a range of stellar ages. Star particles accreted from less dense environments, such as from a filament or a stream from a disrupting satellite in the outskirts of the galaxy, appear as more sparse and uncorrelated events.

\begin{figure*}
\includegraphics[width=\hsize]{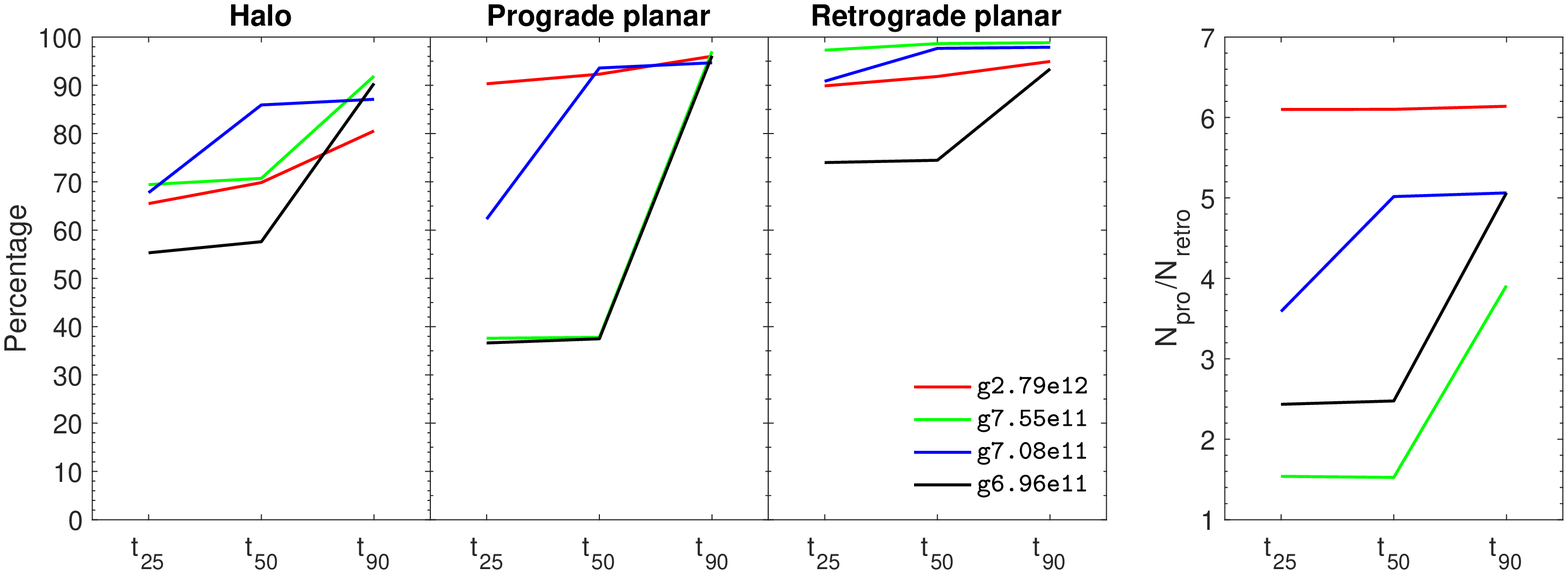}
\caption{Fraction of the halo, prograde planar, and retrograde planar stellar population in the main halo out to a radius of  $50\kpc$  as a function of time, relative to the distribution at the present time. From the first to the third panel:  the percentage of the halo, the prograde planar, and the retrograde planar populations calculated at the time $t_{25}$, $t_{50}$, and $t_{90}$, respectively. Right panel: ratio $N_{\rm pro}/N_{\rm retro}$ between the number of the prograde and retrograde planar star particles  calculated at the time $t_{25}$, $t_{50}$, and $t_{90}$. The legend is the same for all the panels.}\label{percentplot}
\end{figure*}

The percentages of low-metallicity star particles that enter the main halo (out to a radius of $50\kpc$) at times $t_{25}$, $t_{50}$, and $t_{90}$ are summarised in Figure~\ref{percentplot} and the fraction is reported in Table~\ref{tree_table}.  In all simulations, a majority (between 54 and 72 per cent) of all low-metallicity star particles are already brought in by $t_{25}$. This percentage is in line with the overall build-up of mass in their stellar haloes at these radii. The majority of these stars that are brought in by the early accretion events via the galactic building blocks (or are part of the main progenitor) of the proto-MW-like galaxy end up on halo orbits, consistent with the picture of a more metal-poor population that is distributed in a spheroid \citep[e.g.,][]{ElBadry18}. However, as already demonstrated in Sections~\ref{spheroidsec} and~\ref{mwsim} and Figure~\ref{action2}, a non-negligible fraction of star particles are deposited with planar orbits and  prograde motion. 
   
   \begin{table*}
\centering
\caption{Fraction of low-metallicity star particles ($\FeH\leq-2.5$) in the main halo as a function of time. The fractions are reported for all the sample, the halo low-metallicity population, the prograde planar, and the retrograde planar population brought in by the building blocks and merger events at the time $t_{25}$, $t_{50}$, and $t_{90}$. At each time, the ratio between the prograde and the retrograde planar population, $\mathrm{N_{pro}}/\mathrm{N_{retro}}$ is also reported. $t_{25}$, $t_{50}$, and $t_{90}$ are defined as the time when the main halo has assembled the $25$ per cent, the $50$ per cent, and the $90$ per cent of the present stellar mass.}
\label{tree_table}
\resizebox{\textwidth}{!}{
\begin{tabular}{c|c|c|c}
\hline
&  Accreted at $t\leq t_{25}$  & Accreted at   $t\leq t_{50}$  &  Accreted at $t\leq t_{90}$   \\
Simulation & All | Halo | Pro | Retro | $\mathrm{N_{pro}}/\mathrm{N_{retro}}$ &  All | Halo | Pro | Retro | $\mathrm{N_{pro}}/\mathrm{N_{retro}}$   & All | Halo | Pro | Retro | $\mathrm{N_{pro}}/\mathrm{N_{retro}}$   \\
\hline
\texttt{g2.79e12} &  0.72 | 0.66 | 0.90 | 0.90 | 6.25 &  0.76 | 0.70 | 0.92 | 0.92 | 6.25 &  0.84 | 0.81 | 0.96 | 0.95 | 6.25   \\ 
\texttt{g7.55e11} &  0.67 | 0.69 | 0.38 | 0.97 | 1.54 &  0.68 | 0.71 | 0.38 | 0.99 | 1.52 &  0.94 | 0.92 | 0.97 | 0.99 | 3.85   \\ 
\texttt{g7.08e11} &  0.71 | 0.68 | 0.62 | 0.91 | 3.57 &  0.89 | 0.86 | 0.94 | 0.98 | 5.00 &  0.90 | 0.87 | 0.95 | 0.98 | 5.00   \\ 
\texttt{g6.96e11} &  0.54 | 0.55 | 0.37 | 0.74 | 2.44 &  0.56 | 0.58 | 0.38 | 0.75 | 2.50&  0.92 | 0.90 | 0.96 | 0.93 | 5.00   \\ 
 \hline
   \end{tabular}}
   \end{table*}

Figures~\ref{injection}~and~\ref{percentplot} and Table~\ref{tree_table} provide a clear answer regarding the origin of this particular population. The majority of planar low-metallicity stars, both in prograde and retrograde orbits, are brought in at early times, with only a small role for the alternative scenario of later minor mergers. Some building blocks contribute star particles that  end up in prograde motion as well as some that end up in retrograde orbits; this is especially true for merging events in the very chaotic early phases of the build-up of the proto-galaxies. In line with this scenario, \citet{Horta20} detect an observational signature in the chemodynamical properties of the MW bulge, pointing out to an accretion event that happened during the building blocks phase. Because all VMP star particles in the simulations are born within $\sim4\Gyr$ (see Figures~\ref{metage}~and~\ref{injection}), before the formation of the stable, thin and extended disk in these galaxies \citep[see Figure 10 in][]{Buck20}, we can rule out the hypothesis that this VMP population formed at later times in the disk itself. 

However, there are also interesting differences between the galaxies in the populations of star particles that end up on retrograde and prograde planar orbits. For \texttt{g7.55e11} and \texttt{g6.96e11}, $\sim 37$ per cent of the present day prograde planar stars are already in place at $t_{25}$, whereas, for \texttt{g7.08e11} and  \texttt{g2.79e12}, this number is $\sim 62$ and $\sim 90$ per cent, respectively. This difference can be explained by a difference in the formation and accretion history. Simulated galaxies with a more extended merger/accretion history, such as \texttt{g7.55e11} and \texttt{g6.96e11}, will homogeneously gain star particles with prograde planar and halo-like orbits across cosmic time compared to simulations, like \texttt{g7.08e11} and  \texttt{g2.79e12}, that have a very quiet accretion history after the first few Gyr. This is also in agreement with \citet{Gomez17}, where they show with the Auriga simulations that later merger events can bring a significant percentage of old and metal-poor star particles with prograde planar motion into the stellar disks. A very similar result has been found by \citet{Scannapieco11} looking at $\Lambda$CDM simulations, showing that late accretions can deposit their stars in a nearly co-planar orbits.

The population of retrograde planar stars, on the other hand, shows a more consistent picture among the simulations with different accretion histories. The majority of this population ($\gtrsim 90$ percent of this final population) has been assembled at $t_{25}$ in simulations \texttt{g2.79e12}, \texttt{g7.55e11}, and \texttt{g7.08e11}. For galaxy \texttt{g6.96e11} this value is $\sim 74$ percent. 

The picture that emerges here is that whereas retrograde planar low-metallicity stars are almost exclusively tracing a phase of very early build-up, their prograde counterparts are sampling more the full accretion history of the galaxy. This can be explained by the fact that prograde mergers experience the tidal forces of the main galaxy's gravitational potential well for a prolonged period \citep[see e.g.,][]{Abadi03,Penarrubia02}, meaning that later mergers have a much higher chance to sink deep into the potential well if they are prograde rather than retrograde. While retrograde mergers might still happen at later times as well, their higher relative impact speed results in a more violent tidal force, and their stars will typically be disrupted at much larger radii in the Galactic halo. Another reason for this might be given by a simple selection bias. The general suppression of late time retrograde mergers results from the fact that we are looking at disk galaxies to start with. For example, \citet{Martin2018} showed that late time retrograde mergers trigger stronger morphological changes compared to prograde mergers. Thus, by selecting galaxies with a strong stellar disk we might be biasing ourselves towards less retrograde late time mergers. 

\section{Conclusions}\label{conclusions}
We use the NIHAO-UHD cosmological zoom-in simulations, a suite of high-resolution simulated spiral galaxies, with the aim to explore the properties of the oldest and most metal-poor stars, such as their accretion time, their age, and the relation between their kinematical properties and the origin of the star particles. These properties are difficult to infer from observational data, and we use the cosmological simulations as a tool to interpret the observations. In particular, we detect in the NIHAO-UHD simulations the signature of a low-metallicity population that spatially and kinematically resides in the disk. Such ensemble is composed of prograde and retrograde star particles, as also observationally detected by \cite{Sestito19,Sestito20} and \cite{DiMatteo20}. As in the observations, all the simulated galaxies agree on the prevalence for a prograde planar population. While the halos of the simulated galaxies are more significantly rotating that the observed halo stars of the MW, it is also clear that, independently, they also host a population of prograde planar stars that follow the velocity distribution of the more metal-rich thick disk.

We find that the presence of the low-metallicity star particles that kinematically inhabit the disk is explained by two scenarios. The first, dominating, scenario is that during the first few $\Gyr$, the proto-galaxy is undergoing a violent process of assembling, during which the building blocks (and the main MW progenitor) of stellar mass of $10^5-10^9 \msun$ are merging together. During this phase the proto-galaxy and, therefore, the proto-disk are still assembling and the gravitational potential well is much shallower than at the present day. Therefore, the merging building blocks, often with a size comparable to the main MW progenitor (see e.g. Fig.~\ref{tree}), can deposit their star particles in the inner part of the main halo, either in prograde or retrograde planar orbits.  

The second scenario is linked to later merger events. As the proto-galaxy grows in mass and the disk forms, later accretions bring in more prograde planar star particles, but fewer star particles on retrograde planar orbits. This may result from the fact that satellites on prograde orbits tend to sink onto the plane before disrupting, whereas retrograde orbits less so. When those galaxies  finally disrupt they deposit their star particles on prograde planar orbits. Late time retrograde mergers, on the other hand, increase the relative impact speed of merging dwarf and main galaxy leading to more severe tidal forces which violently disrupt the dwarf and deposit star particles on more eccentric halo-like orbits. 

A third possible scenario, the formation of the low-metallicity stars in the disk, has to be ruled out in these simulations. The VMP star particles formed within $\sim4\Gyr$, when the proto-galaxy is still assembling. The formation of the disk and the settling of the ISM happens after these low-metallicity star particles were already born, either in the building blocks, or in the satellites that will later be accreted.

 There is ample evidence for (massive) merger events in the early MW \citep[\eg][]{Belokurov18,Helmi18,Koppelman18,Koppelman19,Barba19,Myeong19,Bonaca20}. The properties of the very metal-poor stars cannot be viewed in isolation from these events as, for instance, evidenced by how well they trace the spatial and kinematical structure of the thick disk in all simulations, and they can help to constrain this picture further. Additionally, we find that, independently of the exact formation history of the galaxy, the vast majority of the star particles on retrograde planar orbits has been deposited there at very early times, therefore this population might provide a unique opportunity to investigate the very early merging phase of the MW.
 
In conclusion, the simulated MW-like galaxies in the NIHAO-UHD suite confirm that the low-metallicity stars are an ideal probe of the galaxy formation during the infant Universe, and therefore, Galactic Palaeontology (or Archaeology) surveys should hunt for the most metal-poor stars, not only in the spheroidal components of the MW, but also in its disk.

\section*{Acknowledgements}
The authors would like to thank Amina Helmi for insightful discussions that helped to improve this manuscript. FS thanks the Initiative dExcellence IdEx from the University of Strasbourg and the Programme Doctoral International PDI for funding his Ph.D. This work has been published under the framework of the IdEx Unistra and benefits from a funding from the state managed by the French National Research Agency as part of the investments for the future program. FS acknowledges the support and funding of the Erasmus+ programme of the European Union. TB acknowledges support by the European Research Council under ERC-CoG grant CRAGSMAN-646955. TB gratefully acknowledges the Gauss Centre for Supercomputing e.V. (\url{https://www.gauss-centre.eu}) for funding this project by providing computing time on the GCS Supercomputer SuperMUC at Leibniz Supercomputing Centre (\url{https://www.lrz.de}). This research was carried out on the High Performance Computing resources at New York University Abu Dhabi; Simulations have been performed on the ISAAC cluster of the Max-Planck-Institut f\"ur Astronomie at the Rechenzentrum in Garching and the DRACO cluster at the Rechenzentrum in Garching. We greatly appreciate the contributions of all these computing allocations. FS and NFM gratefully acknowledge support from the French National Research Agency (ANR) funded project "Pristine" (ANR-18-CE31-0017) along with funding from CNRS/INSU through the Programme National Galaxies et Cosmologie and through the CNRS grant PICS07708. ES gratefully acknowledge funding by the Emmy Noether program from the Deutsche Forschungsgemeinschaft (DFG). AO is funded by the Deutsche
Forschungsgemeinschaft (DFG, German Research Foundation) - MO 2979/1-1. The authors acknowledge the support and funding of the International Space Science Institute (ISSI) for the international team "Pristine". This research made use of the following {\sc{python}} packages: {\sc{pynbody} and {\sc tangos}} \citep{pynbody,tangos}, {\sc{matplotlib}} \citep{matplotlib}, {\sc{SciPy}} \citep{scipy}, {\sc{NumPy}} \citep{numpy}, {\sc{IPython}} and {\sc{Jupyter}} \citep{ipython,jupyter}

\section*{Data Availability}
The data underlying this article will be shared on reasonable request to the corresponding author.


\bibliographystyle{mnras}
\bibliography{Refs} 

\bsp	
\label{lastpage}
\end{document}